\documentclass[12pt]{article}

\def\cosech{\mathop{\rm cosech}\nolimits}
\def\sech{\mathop{\rm sech}\nolimits}

\title{
Conditionally exactly solvable potential and dual transformation in quantum mechanics}
\author{B Bagchi$^{\dagger}$ and C Quesne$^{\ddagger}$\\
$^{\dagger}$ {\small Department of Applied Mathematics, University of Calcutta,}\\
{\small 92 Acharya Prafulla Chandra Road, Calcutta 700009, India}\\
$^{\ddagger}$ {\small Physique Nucl\'eaire Th\'eorique et Physique
Math\'ematique,  Universit\'e Libre de Bruxelles,} \\ 
{\small Campus de la Plaine CP229, Boulevard~du Triomphe, B-1050 Brussels,
Belgium}\\
{\small E-mail: bbagchi123@rediffmail.com and cquesne@ulb.ac.be}}
\date{ }
\begin{document}
\baselineskip=20pt plus 1pt minus 1pt
%%%%%%%%%%%%%%%%%%%%%%%%%%%%%%%%%%%%%%%%%%%%%%%%%%%%%%%%%%
\maketitle
\begin{abstract}
We comment that the conditionally exactly solvable potential of Dutt {\sl et al} (1995 {\sl J.\ Phys.\ A:
Math.\ Gen.} {\bf 28} L107) and the exactly solvable potential from which it is derived form a dual
system.
\end{abstract}

\vspace{0.5cm}


\newpage

Conditionally exactly solvable (CES) potentials have received considerable attention in recent
times~\cite{dutra, dutt, grosche, znojil, roychoudhury}. The main feature of such potentials is that one
or more coupling constants in them are fixed to a specific value. There have been instances of CES
potentials running into inconsistencies with threshold boundary conditions~\cite{grosche,
znojil}, but there do exist some which possess valid asymptotic behaviour. One such acceptable class is
the one proposed by Dutt {\sl et al\/}~\cite{dutt} sometime ago and which reads~\footnote{There is
also another class of CES potentials proposed in~\cite{dutt} but by a redefinition of
parameters~\cite{roychoudhury} it can be made equivalent to (\ref{eq:V}) and so is not considered
here.}
\begin{equation}
  V(y) = \frac{A}{1 + e^{-2y}} - \frac{B}{(1 + e^{-2y})^{1/2}} - \frac{3}{4(1 + e^{-2y})^2}
  \label{eq:V} 
\end{equation}
where $y \in (-\infty, \infty)$ and $A$, $B$ are some real parameters defining the shape of the
potential. Note the presence of the fixed numerical value $-\frac{3}{4}$ for one of the coupling
constants in $V(y)$ that provides its identification as a CES. Potential (\ref{eq:V}) has for its associated
eigenfunctions
\begin{equation}
  \psi_n(y) = z^{\frac{1}{4}} (z-1)^{-\left(\frac{c}{2} - \frac{B}{4c}\right)} (z+1)^{-\left(\frac{c}{2} -
  \frac{B}{4c}\right)} P_n^{\left(\frac{B}{2c} - c, - \frac{B}{2c} - c\right)}(z),  \label{eq:w-f} 
\end{equation}
where $z = 1 + e^{-2y}$ and $c$ is related to the energy eigenvalue $-\epsilon_n$ as $c = n +
\frac{1}{2} + \sqrt{\epsilon_n}$. Actually $\sqrt{\epsilon_n}$ satisfies a complicated cubic equation
but it has been observed~\cite{roychoudhury} that only one of its roots is compatible with the
normalizability condition.\par
%
%-----------------------------------------------------------------------------------------------------------
%  
Some remarks are in order concerning the derivation of the eigenfunctions (\ref{eq:w-f}) and the energy
eigenvalue equation. The Schr\"odinger equation is subjected to a coordinate transformation and the
transformation function is chosen in such a manner that corresponding to an exactly solvable (ES)
potential one has a new analytically solvable one. It turns out that for a half-line-full-line mapping
function, one can generate the CES potential for some known shape-invariant potential as an input. The
available energy eigenvalues and eigenfunctions of the latter then furnish the corresponding ones for
the former.\par
%
%-------------------------------------------------------------------------------------------------------
%
The purpose of this comment is to establish that the CES potential (\ref{eq:V}) and its accompanying
shape-invariant ES potential are actually dual partners in the sense that the corresponding
time-independent Schr\"odinger equations are mapped to each other under appropriate space
transformations called the dual transformations~\cite{mittag98, mittag92, grant}. The latter are known
to relate some apparently unconnected problems both in classical and quantum mechanics. The
one-dimensional harmonic oscillator and the Coulomb problem~\cite{bateman}, the latter and the
isotropic oscillator~\cite{kostelecky, lahiri}, the P\"oschl-Teller and infinite potential well
problems~\cite{mittag98} are some examples of dual systems (DS).\par
%
%------------------------------------------------------------------------------------------------------
%
In the present context, let us write down the following set of DS given by the pair of Schr\"odinger
equations (with $\hbar = 2m = 1$)
\begin{eqnarray}
  && \left[- \frac{d^2}{dx^2} + \lambda \left(\frac{dy}{dx}\right)^2 + \nu \frac{dy}{dx}\right] \psi =
        \mu \psi  \label{eq:DS-1} \\
  && \left[- \frac{d^2}{dy^2} - \frac{1}{2} \{x, y\} - \mu \left(\frac{dx}{dy}\right)^2 + \nu
        \frac{dx}{dy}\right] \phi = - \lambda \phi  \label{eq:DS-2} 
\end{eqnarray}
where $\mu$ and $-\lambda$ are the energies, $\nu$ is a constant, $\{x, y\}$ is the Schwarzian
derivative, which can be written as
\begin{equation}
  \{x, y\} = - \frac{1}{y^{\prime2}} \left[\frac{d}{dx} \left(\frac{y''}{y'}\right) - \frac{1}{2}
  \left(\frac{y''}{y'}\right)^2\right]  \label{eq:Schwarz}
\end{equation}
the primes denoting derivatives with respect to the variable $x$. The wave functions $\psi$ and $\phi$
are related in the manner
\begin{equation}
  \psi = \left(\frac{dx}{dy}\right)^{1/2} \phi.  \label{eq:wf-transform} 
\end{equation}
In DS as above, the energy and the coupling constant $\mu$ and $\lambda$ exchange roles. Further DS
are meaningful when the potential and its partner are expressible as
\begin{eqnarray}
  && W(x) = \lambda \left(\frac{dy}{dx}\right)^2 + \nu \frac{dy}{dx}  \label{eq:W} \\
  && U(y) = - \mu \left(\frac{dx}{dy}\right)^2 + \nu \frac{dx}{dy} - \frac{1}{2} \{x, y\}.  \label{eq:U} 
\end{eqnarray}
Generalizations to include integral or even fractional powers in the derivatives in (\ref{eq:W}) and
(\ref{eq:U}) are straightforward.\par
%
%-------------------------------------------------------------------------------------------------
%
We now set $dy/dx = \coth x$, i.e., $y = \log\sinh x$: in other words, for $y \in (-\infty, \infty)$ the
variable $x \in (0, \infty)$. It implies from (\ref{eq:W}) that the Schr\"odinger equation (\ref{eq:DS-1})
has the potential $W(x) = \tilde{W}(x) + \alpha(\alpha-1)$, where
\begin{equation}
  \tilde{W}(x) = - 2\beta \coth x + \alpha(\alpha-1) \cosech^2x \qquad x \in (0, \infty)
\end{equation}
which is ES (for $\beta > \alpha^2$, $\alpha>0$) and shape invariant as well. The energy eigenvalues 
are given by
\cite{cooper}
\begin{equation}
  E_n = - \left(\frac{\beta}{\alpha+n}\right)^2 - (\alpha+n)^2 \qquad n=0, 1, 2, \ldots.  \label{eq:E}
\end{equation}
Indeed the correspondence with (\ref{eq:DS-1}) is provided by the following identifications
\begin{eqnarray}
  \lambda & = & \alpha(\alpha-1)  \nonumber \\
  \nu & = & - 2\beta \label{eq:identification}\\
  \mu & = & E_n + \alpha(\alpha-1)  \nonumber.
\end{eqnarray}
\par
%
%------------------------------------------------------------------------------------------------
% 
We next enquire into the dual potential $U(y)$. It is easy to work out $\{x, y\}$ from
(\ref{eq:Schwarz}) as
\begin{equation}
  \{x, y\} = - \left(\sech^2x \tanh^2x + \sech^2x - \frac{1}{2} \sech^4x\right). 
\end{equation}
Hence we find from (\ref{eq:U}) and (\ref{eq:DS-2}) that $U(y) = \tilde{U}(y) + \frac{1}{4}$, where
\begin{eqnarray}
  && \tilde{U}(y) = \left[\left(\frac{1}{2} - \mu\right) \tanh^2x + \nu \tanh x - \frac{3}{4} \tanh^4x
           \right]_{x = \sinh^{-1}(e^y)} \label{eq:Utilde}\\
  && \epsilon_n = \alpha(\alpha-1) + \frac{1}{4}.  \label{eq:epsilon}
\end{eqnarray}
The relation between the energies $- \epsilon_n$ and $E_n$ turns out to be
\begin{equation}
  \epsilon_n + E_n = \mu + \frac{1}{4}.
\end{equation}
On elimination of the parameter $\alpha$, from (\ref{eq:E}), (\ref{eq:identification}) and
(\ref{eq:epsilon}) we then get a cubic equation in $\sqrt{\epsilon_n}$, similar to that given
in~\cite{dutt}. Further, in terms of the variable $y \in (-\infty, \infty)$, $\tilde{U}(y)$ in
(\ref{eq:Utilde}) translates to
\begin{equation}
  \tilde{U}(y) = \frac{\frac{1}{2} - \mu}{1 + e^{-2y}} + \frac{\nu}{(1 + e^{-2y})^{1/2}} -
  \frac{3}{4(1 + e^{-2y})^2} \qquad y \in (-\infty, \infty)
\end{equation}
which is identical to the potential in (\ref{eq:V}) for $\mu = \frac{1}{2} - A$ and $\nu = - B$.
Finally, the wave functions for $\tilde{U}(y)$ can be obtained from (\ref{eq:wf-transform}) and yield
the same form as in (\ref{eq:w-f}).\par
%
%------------------------------------------------------------------------------------------------
%
To conclude we have demonstrated that the CES potential of Dutt {\sl et al\/} and its ES partner form a
DS.\par
%
%----------------------------------------------------------------------------------------------------------------
%
One of us (BB) gratefully acknowledges the support of the National Fund for Scientific Research (FNRS),
Belgium, and the warm hospitality at PNTPM, Universit\'e Libre de Bruxelles, where this work was carried
out. CQ is a Research Director of the National Fund for Scientific Research (FNRS), Belgium.\par
%
%=========================================================
%

\end{document}